\documentclass[aps,twocolumn,prl,floatfix,superscriptaddress,showpacs]{revtex4-1}
\usepackage{epsf}
\usepackage{epsfig}
\usepackage{graphicx}
\usepackage{dcolumn}
\usepackage{braket}
\usepackage{bm}
\usepackage{amsfonts}
\usepackage{amsmath}
\usepackage{amssymb}
\usepackage{color,soul}

\newcommand{\bea}{\begin{eqnarray}}
\newcommand{\eea}{\end{eqnarray}}

\newcommand{\ci}{\mathrm{i}}

\begin{document}

\title{Floquet chiral edge states in graphene}
\author{P. M. Perez-Piskunow}
\affiliation{Instituto de F\'{\i}sica Enrique Gaviola (CONICET) and FaMAF, Universidad Nacional de C\'ordoba, Argentina}
\author{Gonzalo Usaj}
\affiliation{Centro At{\'{o}}mico Bariloche and Instituto Balseiro,
Comisi\'on Nacional de Energ\'{\i}a At\'omica, 8400 Bariloche, Argentina}
\affiliation{Consejo Nacional de Investigaciones Cient\'{\i}ficas y T\'ecnicas (CONICET), Argentina}
\author{C. A. Balseiro}
\affiliation{Centro At{\'{o}}mico Bariloche and Instituto Balseiro,
Comisi\'on Nacional de Energ\'{\i}a At\'omica, 8400 Bariloche, Argentina}
\affiliation{Consejo Nacional de Investigaciones Cient\'{\i}ficas y T\'ecnicas (CONICET), Argentina}
\author{L. E. F. Foa Torres}
\affiliation{Instituto de F\'{\i}sica Enrique Gaviola (CONICET) and FaMAF, Universidad Nacional de C\'ordoba, Argentina}

\begin{abstract}
We report on the emergence of laser-induced chiral edge states in graphene ribbons. Insights on the nature of these Floquet states is provided by an analytical solution which is complemented with numerical simulations of the transport properties. Guided by these results we show that graphene can be used for realizing non-equilibrium topological states with striking tunability: While the laser intensity can be used to control their velocity and decay length, changing the laser polarization switches their propagation direction.  

\end{abstract}
\pacs{73.22.Pr; 73.20.At; 72.80.Vp; 78.67.-n}
\date{\today}
\maketitle

\textit{Introduction.--} 
Topological insulators (TI) \cite{Fu2007,Hasan2010} are an e\-xo\-tic family of materials \cite{Koenig2007} where a bulk gap is bridged by edge states which propagate even in the presence of disorder, thereby providing a potentially outstanding platform for quantum computation \cite{Moore2010,Nayak2008} or spintronics \cite{Zutic2004}, among many  other applications \cite{Hasan2010}. 
Graphene \cite{Novoselov2005a,Zhang2005,Geim2009}, on the other hand, has emerged as a novel material with record electronic \cite{CastroNeto2009}, thermal \cite{Balandin2008}, mechanical and optical properties \cite{Bonaccorso2010,Xia2009,Ren2012,Tielrooij2013}. Endowing graphene with protected edge states would unite the best of both materials.

Since graphene is a zero gap semiconductor a very first step is the creation of a bulk band-gap. Predictions indicate that a circularly polarized laser can do this task \cite{Oka2009,Lopez-Rodriguez2008,Kibis2010,Calvo2011,Savelev2011}---this was verified by a recent experiment at the surface of a TI \cite{Wang2013}. Although laser induced band-gaps appear both at the Dirac point and away from it, the most promising ones are the latter \cite{notes1}, also called \textit{dynamical gaps} \cite{Syzranov2008}, which occur at half the photon energy ($\hbar\Omega$) above/below the Dirac point and can be reached within an experimentally relevant set of parameters \cite{Calvo2011} ($n\sim 2.5\times 10^{11}$ cm$^{-2}$ for $\hbar\Omega=100$meV, $\lambda\sim 10 \mu$m). Once the bulk gap opens, one should look for  Floquet edge states (FES). Such intriguing states were proposed in \cite{Lindner2011,Kitagawa2010,other} and realized  recently in photonic crystals \cite{Rechtsman2013} but experiments were not reported in condensed matter so far. 
Crafting FES within dynamical gaps in graphene would extend the realm of \textit{Floquet topological insulators} (FTI) \cite{Lindner2011,Cayssol2013} and lead to a new  playground for optoelectronics \cite{Bonaccorso2010,Glazov2013}.

Here we show how  chiral edge states emerge at the dynamical bandgaps in graphene. To such end we use Floquet theory (FT) and combine numerics with an explicit analytical solution for the edge states. Our analysis reveals that these states decay exponentially towards the bulk with a decay length that depends only on the ratio of the field's frequency and its intensity. More importantly, these FES turn out to be chiral, \textit{i.e.} all the states on each edge of the sample propagate in the \textit{same} direction, like in one-way streets.
Additional simulations of charge transport confirm the edge states' robustness against disorder, highlighting their chirality.

\textit{Model for irradiated graphene and Floquet theory.--} To start with let us introduce our model Hamiltonian for irradiated graphene. By using Weyl's gauge the electromagnetic field (incident perpendicularly to the graphene sheet) is modeled through a vector potential $\bm{A}(t)=\Re\left\{\bm{A}_0e^{\mathrm{i}\Omega t}\right\}$. We take $\bm{A}_{0}=A_{0} (\hat{\bm{x}}+\ci\hat{\bm{y}})$ that yields a circularly polarized field. The interaction with the laser is modeled through the Hamiltonian
\begin{equation}
\hat{\cal{H}}(t)=v_F\,\bm{\sigma}\cdot\left[\bm{p}+
\frac{e}{c}\bm{A}(t)\right], 
\label{eq_laser-graphene-Hamiltonian} 
\end{equation}
where $v_{F}\simeq 10^{6} m/s$ denotes the Fermi velocity, $\bm{\sigma}=(\sigma_x,\sigma_y)$ the Pauli matrices describing the pseudo-spin degree of freedom---the real spin does not play a role here.

Given the time-periodicity of the Hamiltonian, we resort to FT \cite{Shirley1965,Grifoni1998}. The main advantage is that it provides a non-perturbative solution, as required in our case. The solutions of the time-dependent Schr\"odinger equation are of the form $\psi_{\alpha}(r,t)=\exp(-\ci\varepsilon_{\alpha}t/\hbar) \phi_{\alpha}(r,t)$ where $\phi_{\alpha}(t)$ have the same time-periodicity as the Hamiltonian, $\phi_{\alpha}(t+T)=\phi_{\alpha}(t)$ with $T=2\pi/\Omega$, and is the solution of $\hat{\cal{H}}_F\phi_{\alpha}(t) =\varepsilon_{\alpha}\phi_{\alpha}(t)$, where $\hat{\cal{H}}_F=\hat{\cal{H}}-i\hbar \frac{\partial}{\partial t}$ is the Floquet Hamiltonian and $\varepsilon_{\alpha}$ determine the quasi-energy spectrum. 
This is essentially an eigenvalue problem in the space $\cal{R}\otimes \cal{T}$, where $\cal{R}$ is the usual Hilbert space and $\cal{T}$ is the space of periodic functions with period $T$ (spanned by the set of orthonormal functions $\exp(\ci m\Omega t)$, where $m$ is an integer that labels the Floquet replicas). 
Written in this basis, $\hat{\cal{H}}_F $ is a time-independent infinite matrix where only the replicas with $\Delta m=\pm1$ are coupled by the radiation field. 
The solutions can be obtained by truncating the Floquet basis, keeping only the replicas with $|m|\leq N_\mathrm{max}$  and choosing $N_\mathrm{max}$ so that desired magnitudes converge.  
Then one can compute the average density of states, the dc conductance, and all physical quantities of interest \cite{Shirley1965,Kohler2005}.
Whereas the Dirac Hamiltonian (\ref{eq_laser-graphene-Hamiltonian}) is suitable for analytic calculations at low energies, in some cases, as for transport calculations, numerically solving the more complete tight-binding model is more convenient \cite{Calvo2011}. Here we use both methods.

\textit{Unveiling the emergence of a Floquet topological phase.--}
Thanks to graphene's peculiar electronic structure, a laser can induce a \textit{resonant} coupling between electronic states for any frequency in the range of interest ($\hbar\Omega\lesssim150$meV). Indeed, there is always a state below the Dirac point that can be coupled to a corresponding state above the Dirac point with the excess energy being released or absorbed by the laser field. When looked from the viewpoint of FT, these two states are degenerate (belonging to different replicas) and are split by the laser field. This leads to the possibility of creating a bulk bandgap through a laser excitation as described in \cite{Oka2009}, reaching experimentally accessible magnitudes in the mid-infrared range \cite{Calvo2011} (as recently observed \cite{Wang2013}). In the following we explicitly show that these \textit{bulk} bandgaps are accompanied by chiral FES \cite{Rudner2013}.

\begin{figure}[tb]
\includegraphics[width=\columnwidth]{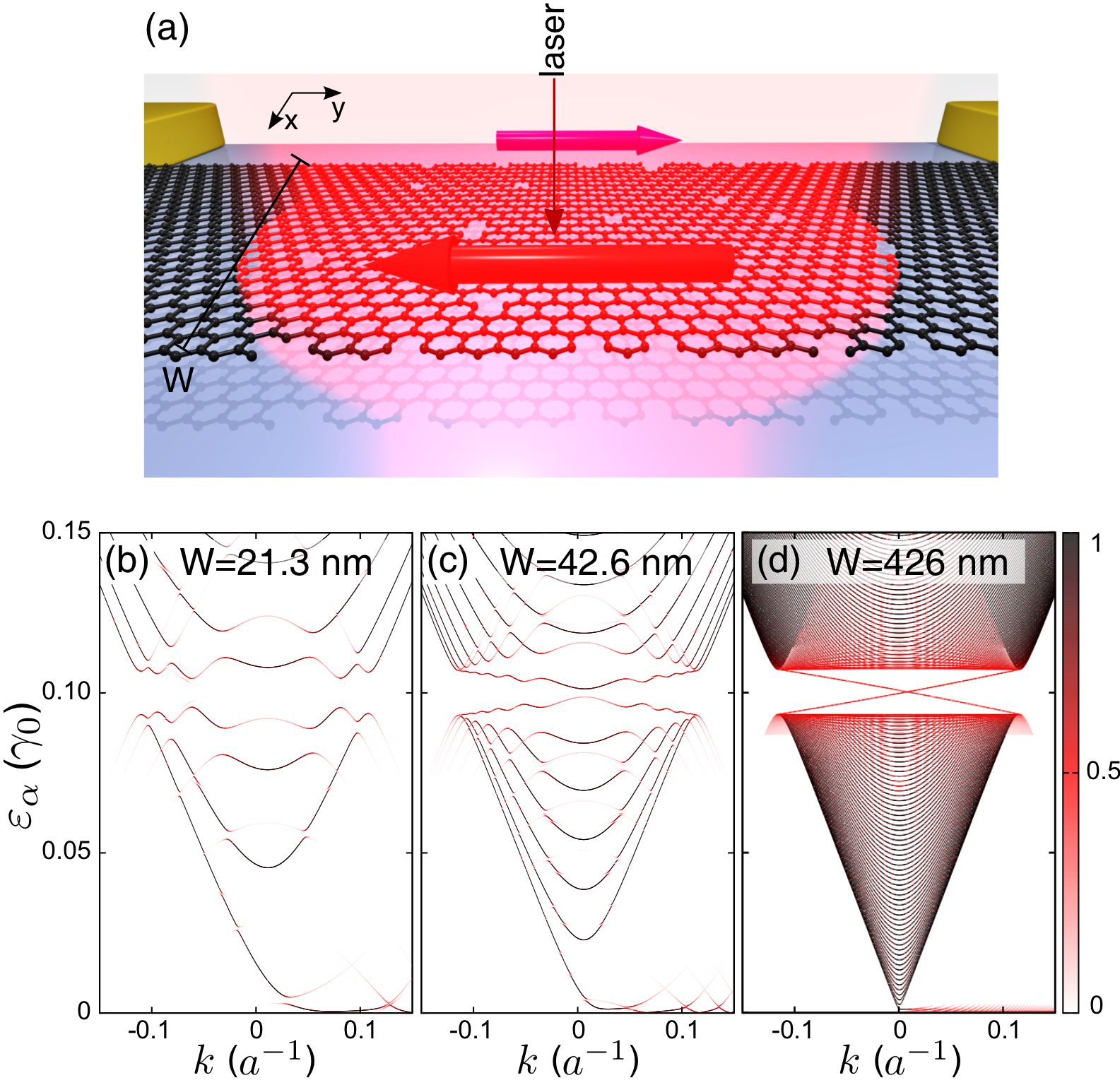}
\caption{(Color online)
(a) Scheme of the setup: a laser of frequency $\Omega$ illuminates a section of an eventually disordered graphene ribbon. The circular polarization is shown to induce chiral edge states 
(b)-(d) Typical evolution of the quasienergy dispersion as the system width $W$ is increased ((b) $W=21.3$nm, (c) $W=42.6$nm and (d) $W=426$nm) obtained from a numerical solution of  (\ref{eq_laser-graphene-Hamiltonian}). 
The laser frequency in this plot is $\hbar\Omega=0.2\gamma_0$ and $N_\mathrm{max}=2$. The color indicates the weight contributing to the average density of states. }
\label{fig1}
\end{figure}

Figures \ref{fig1}(b)-(d) show the quasi-energy dispersion of zigzag ribbons, described with Hamiltonian (\ref{eq_laser-graphene-Hamiltonian}), in the presence of a circularly polarized laser.
Among the many states in the Floquet spectrum only those with a weight on a given channel, say the $m=0$ channel, determine the gap in the time-averaged density of states.
This weight is shown in color scale in Fig. \ref{fig1}(b)-(d). 
As the ribbon width increases one can follow the emergence of states bridging the bulk bandgap.

\begin{figure*}[tb]
\includegraphics[width=0.90\textwidth]{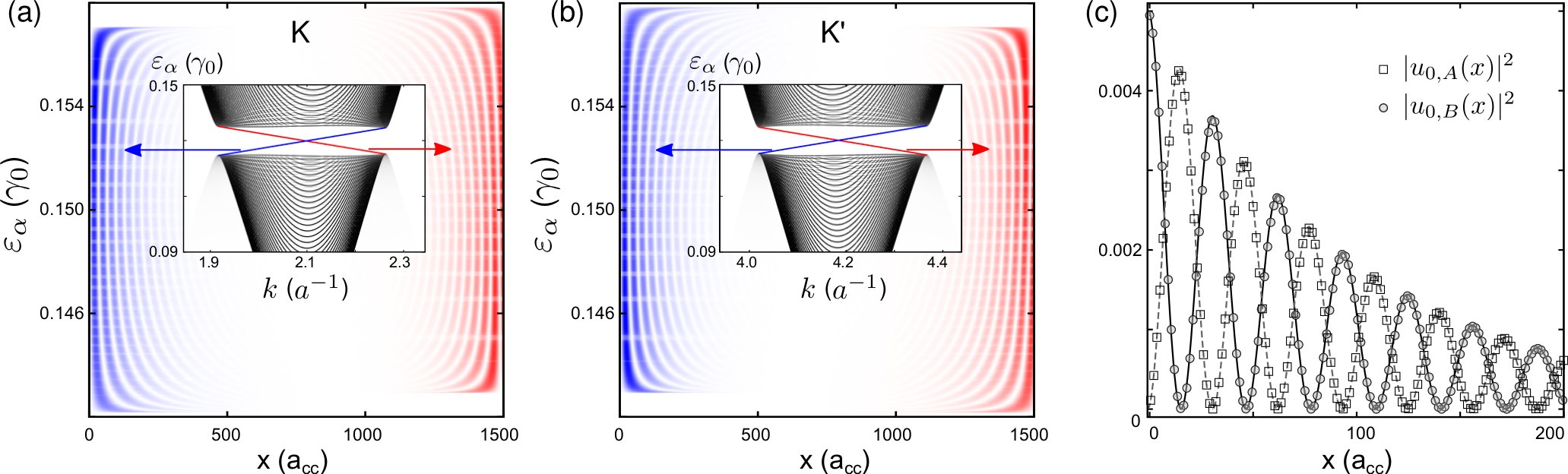}
\caption{(Color online)
Quasi-energy dispersion versus $k$ in the neighborhood of the $K$ ((a) inset) and $K'$ ((b) inset) points. The weight on the $m=0$ Floquet channel is shown in color scale from white (zero) to black (one). The FES bridging the gap at $\hbar\Omega/2$ which have a weight close to $0.5$ are highlighted with red and blue. The spatial distribution of the probability density for those states at each valley (for the $A$ sites) is shown in color scale in the respective main frames. Panel (c) shows a cut along the map shown in (a) for $\varepsilon_{\alpha}=0.15\gamma_0$, this time including type A and B sites (squares and circles) as well as the result from the analytical calculation (dashed and solid lines). The results correspond to circularly polarized light with $\hbar\Omega=0.3\gamma_0$ and $\eta=0.05$ and a zigzag graphene ribbon with $W=213$nm.}
\label{fig2}
\end{figure*}

Figures \ref{fig2}(a)-(b) show that these states are localized on the edges of the graphene ribbon and, more importantly, that each band (shown with red and blue colors) corresponds to states localized on opposite edges of the sample. 
Besides the exponential decay towards the bulk, they also present an oscillatory component as seen in Fig. \ref{fig2}.
In contrast to the usual edge states found in zigzag nanoribbons \cite{Brey2006,CastroNeto2009,Usaj2009}, these FES are away from the Dirac point and are topologically protected as we discuss below. They are present for any ribbon termination (we checked this numerically for a few cases, not shown). 

While Fig.\ref{fig1}(b)-(d) was calculated for a single Dirac cone, Figs. \ref{fig2}(a)-(b) confirms that each Dirac cone (at $K$ and $K'$) supports similar states. The states propagating along a given border have the same sign of $\partial \varepsilon_{\alpha}/\partial k$ whether they come from around $K$ or $K'$. Therefore on each edge they propagate in the \textit{same} direction.

\textit{Analytical solution for the FES in graphene.--} What drives the transition in Fig. \ref{fig1}(b)-(d), and what is the nature of the emerging states? To rationalize this behavior we solve for the eigenstates of the Floquet Hamiltonian for zigzag ribbons in the limit of low laser power and large ribbon width and obtain explicit analytic solutions for the FES. The details will be presented elsewhere, we now give the main results. 


Since radiation does not couple the two inequivalent Dirac points ($K$ and $K'$), we can solve for each of them separately.  By restricting to the subspaces with $m=0$ and $m=1$, the dominant channels near the dynamical gap, the eigenstates of the Floquet equation $\mathcal{H}_F \Phi=\varepsilon \Phi$ near, say, $K$ involve the four-component wave function $\Phi(\bm{r})=\{u_{1A}(\bm{r}),u_{1B}(\bm{r}),u_{0A}(\bm{r}),u_{0B}(\bm{r})\}$, here $0$ and $1$ stand for the Floquet channel index and $A$ and $B$ for the inequivalent lattice sites---the full time dependent solution is $\psi(\bm{r},t)=\exp(-\ci\varepsilon t/\hbar)\phi(\bm{r},t)=\exp(-\ci\varepsilon t/\hbar)\sum_{m,i}\exp(\ci m\Omega t)u_{mi}(\bm{r})$. The boundary conditions imply that the wavefunction has to vanish at the sample edges $u_{mB}(x=0)=0$ and $u_{mA}(x=W)=0$. 
As we look for solutions localized on the edges of the sample we do not need to impose both conditions simultaneously in the wide sample limit (defined below) but only one of them. 
We find that the components of these eigenstates have the simple form $u_{mi}(\bm{r})=\mathcal{C}\, \mathrm{e}^{\ci k_yy} \mathrm{e}^{qx}\mathcal{S}_{mi}(x)$---here $\mathcal{C}$ is a normalization constant and $\mathcal{S}_{mi}(x)$ a trigonometic function. Therefore, the solutions are propagating along the ribbon direction ($y$) and decay as they penetrate into the sample.

The quasi-energy dispersion of the edge states near the center of the dynamical gap can be approximated by
\begin{equation}
\frac{\varepsilon}{\hbar\Omega}\approx \frac{(1+2\eta^2)}{2(1+\eta^2)}\pm\frac{\eta}{2(1+\eta^2)}\frac{k_y}{k_0},
\label{disp1}
\end{equation}
where the adimensional parameter $\eta=e v_F A_0/c\hbar\Omega$ is a measure of the strength of the electron-photon coupling and $\hbar v_Fk_0=\hbar\Omega/2$---the gap is defined by $|\varepsilon-\hbar\Omega/2|\leq\hbar\Omega \eta/2\sqrt{1+\eta^2}$. The expressions for the eigenstates become simpler at the center of the dynamical gap ($\varepsilon = \hbar\Omega/2$). In this case, the solutions vanishing at $u_{mB}(x=0)=0$ are
\begin{eqnarray}
\label{A}
&&\Phi_{K,-}(\bm{r})=\mathrm{e}^{-\ci y/2\xi}\, \mathrm{e}^{-x/2\xi} \sqrt{\frac{1}{2 \xi L_y}}\\
\nonumber
&& \left\{-\cos{k_0 x} + 2\eta\sin{k_0 x}, \ci\sin{k_0 x}, \ci \cos{k_0 x}, -\sin{k_0 x}\right\},
\end{eqnarray}
while the ones vanishing at the opposite edge $u_{mA}(x=W)=0$ have the form
\begin{eqnarray}
\label{B}
&&\Phi_{K,+}(\bm{r})= \mathrm{e}^{-\ci y/2\xi}\, \mathrm{e}^{\tilde{x}/2\xi} \sqrt{\frac{1}{2 \xi L_y}}\\
\nonumber
&&\left\{\ci \sin{k_0 \tilde{x}}, - \cos{k_0 \tilde{x}}, \sin{k_0 \tilde{x}}, -\cos{k_0 \tilde{x}} - 2\eta\sin{k_0 \tilde{x}}\right\}.
\end{eqnarray}
Here $\tilde{x}=x-W$ and $L_y$ is the length of the sample. The oscillation frequency set by the parameter $k_0$ depends on the laser frequency and on the graphene's Fermi velocity. For a mid-infrared laser with $\hbar\Omega\sim 100$meV this leads to an spatial modulation with period of $\sim2$nm. 

Strikingly, the decay rate is governed by the length scale $\xi=\hbar\Omega/eE_0$, which is \textit{independent} of graphene's microscopic parameters, being of the order of $50$nm for a laser of $5$mW/$\mu$m$^2$---here $E_0=\Omega A_0/c$ is the amplitude of the radiation field. 
This result is consistent with the fact that $\xi\sim \hbar v_F/\Delta$ where $\Delta\sim e v_F A_0/c$ is the size of the bulk quasi-energy gap.

We notice that the edge states require a ribbon width $W\gg\xi$ to develop, otherwise they couple to each other and split (as in Figs. \ref{fig1}(b) and (c)). Fig. \ref{fig2}(c) shows the excellent agreement between the numerical results (squares and circles) and the analytical solution (dashed and solid lines).

It follows from Eq. (\ref{disp1}) that the average velocity of the edge states along the ribbon's edge is given by
\begin{equation}
v_\mathrm{ES}=\pm\frac{\eta}{1+\eta^2}v_F
\end{equation}
where the plus (minus) sign corresponds to Eq. (\ref{A}) (Eq. (\ref{B})) \cite{notes2}. Thus, the FES  (of the K cone) located at the opposite sides of the sample have opposite velocities. 

Interestingly enough, similar results are obtained from the analytical solutions corresponding to the $K'$ point as can be anticipated from Fig. \ref{fig2}(a)-(b) (inset). The important result is that the edge states coming from \textit{both} Dirac cones propagate in the \textit{same} direction on a given side of the sample. That is, these FES are chiral and topologically protected and they are expected to exhibit a Hall response \cite{Oka2009}. 
Since  the propagation direction is set by the circularly polarized laser, it can be reversed by changing from, say, left to right handed polarization and so will do the Hall response. 

The presence of two chiral modes is also supported by the calculation of the Chern numbers associated with  $\mathcal{H}_F$  keeping the  $m=0,1$ replicas \cite{Rudner2013}.  Including other replicas introduces additional edge modes (increasing the value of the Chern invariant) but those, having a negligible weight on the $m=0$ subspace,  do not contribute to the dc properties. For elliptical polarization, the results remain qualitatively the same.

\begin{figure}[tbp]
\includegraphics[width=\columnwidth]{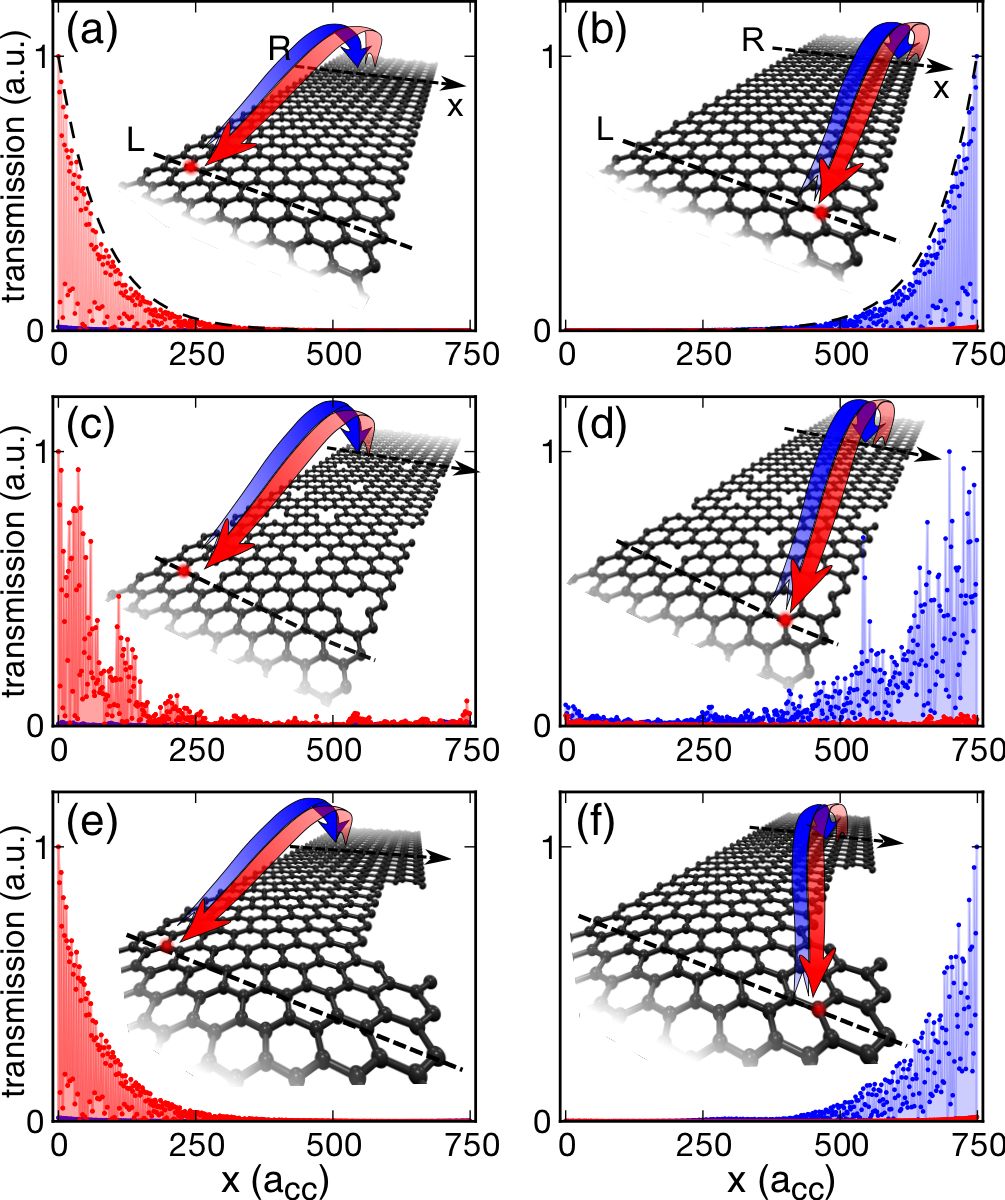}
\caption{(Color online)
Spatial distribution of the total transmission probabilities between point probes located sites $s_1$ on layer L and $s_2$ on layer R. These layers are located $63$ nm from each other and are marked with dashed lines in the insets. $s_1$ is taken as the second atom  counting from the left  (panels (a), (c) and (e)) or from the right (panels (b), (d) and (f)) and are marked with a red dot on the insets. The $x$ coordinate of $s_2$ on layer R is the horizontal axis in these plots ($a_{cc}$ is the carbon-carbon distance). The transmission probabilities $T_{s_1\rightarrow s_2}$ and $T_{s_2\rightarrow s_1}$ are shown as blue and red points respectively and are \textit{normalized} to their maximum value. Panels (a) and (b) are for a pristine sample ($106$nm wide) irradiated with circular polarization ($\hbar\Omega =\gamma_0$, $\eta=3/16$) and with a Fermi energy close to the center of the dynamical gap ($\varepsilon=0.497\times\hbar\Omega$). The black dashed lines are exponentials with a decay rate of $1/\xi$, as indicated by the analytical solution. In panels (c) and (d) random vacancies at 0.1\% are introduced between layers L and R, while in (e) and (f) a strip $2$nm wide and $15.7$nm long was cut from the right side of the sample.}
\label{fig3}
\end{figure}

\textit{Chirality and robustness against disorder.} We have presented both an analytical solution and numerical results confirming the emergence of FES within the dynamical gaps in graphene. But can we numerically test the chirality of these states? For that we use a setup involving transport through local probes: We include weakly coupled probes to two different sites of the irradiated sample ($s_1, s_2$) separated by a distance $d$ along the $y$ direction and compute the total transmission probability between those points \cite{Kohler2005,note-transport}.  
If the states are chiral we expect a strong directional asymmetry between the transmission probability from left to right $T_{s_1\rightarrow s_2}$ and viceversa $T_{s_2\rightarrow s_1}$---blue and red curves, respectively, in Fig. \ref{fig3}.
Panels \ref{fig3}(a) and \ref{fig3}(b), that correspond to pristine graphene, show that there is a marked directional asymmetry in the propagation of the Floquet states, a hallmark of their chiral nature.

Panels \ref{fig3}(c) and \ref{fig3}(d) test the robustness against vacancies included at random in between layers L and R with a density of $0.1\%$. Although the defects produce a small amount of edge to edge scattering, the states remain chiral to a high extent. The same disorder configuration leads to about $40\%$ of the electrons to be backscattered in the absence of radiation. In \ref{fig3}(e) and \ref{fig3}(f) a strip $2$nm wide and $15.7$nm long was cut from the right side of the sample in-between L and R as shown in the insets. Once more, the spatial distribution of the transmission is not being compromised. This supports the conclusion that FES behave as one-way channels where transport is robust against defects.

\textit{Final remarks.--} We have shown how chiral FES emerge inside the dynamical gaps created by the same laser field in graphene. Our analytical expressions for the  FES demonstrate that the decay lenght of these peculiar states is governed only by the applied circularly polarized laser. Further simulations confirm their chirality and highlight their robustness to structural disorder.

Our results indicate that a realization of such topological phase is, though experimentally challenging, feasible in graphene. For a CO$_2$ laser with $\lambda\sim9.3\mu$m ($\hbar\Omega\sim 133$meV), at temperatures of about $4$K a bulk bandgap in excess of $k_B T$ can be achieved for a laser power of about $5$ mW/$\mu m^2$. Furthermore, for a mid-infrared laser the energy shift required to reach the dynamical gaps is in the order of $50-70$meV, a value that can be reached with a standard gate applied to the sample ($n\sim 3\times 10^{11}$ cm$^{-2}$). 
Our results may point a way towards novel devices with great tunability and robustness against disorder.

\noindent
\textit{Acknowlegdments.--} We acknowledge financial support from PICTs 2006-483, 2008-2236, 2011-1552 and
Bicentenario 2010-1060 from ANPCyT and PIP 11220080101821 from CONICET. LEFFT acknowledges support from Trieste's ICTP. We thank Hern\'an  Calvo, Eric Suarez-Morell and Stephan Roche for inspiring discussions. \\


\begin{thebibliography}{40}%
\makeatletter
\providecommand \@ifxundefined [1]{%
 \@ifx{#1\undefined}
}%
\providecommand \@ifnum [1]{%
 \ifnum #1\expandafter \@firstoftwo
 \else \expandafter \@secondoftwo
 \fi
}%
\providecommand \@ifx [1]{%
 \ifx #1\expandafter \@firstoftwo
 \else \expandafter \@secondoftwo
 \fi
}%
\providecommand \natexlab [1]{#1}%
\providecommand \enquote  [1]{``#1''}%
\providecommand \bibnamefont  [1]{#1}%
\providecommand \bibfnamefont [1]{#1}%
\providecommand \citenamefont [1]{#1}%
\providecommand \href@noop [0]{\@secondoftwo}%
\providecommand \href [0]{\begingroup \@sanitize@url \@href}%
\providecommand \@href[1]{\@@startlink{#1}\@@href}%
\providecommand \@@href[1]{\endgroup#1\@@endlink}%
\providecommand \@sanitize@url [0]{\catcode `\\12\catcode `\$12\catcode
  `\&12\catcode `\#12\catcode `\^12\catcode `\_12\catcode `\%12\relax}%
\providecommand \@@startlink[1]{}%
\providecommand \@@endlink[0]{}%
\providecommand \url  [0]{\begingroup\@sanitize@url \@url }%
\providecommand \@url [1]{\endgroup\@href {#1}{\urlprefix }}%
\providecommand \urlprefix  [0]{URL }%
\providecommand \Eprint [0]{\href }%
\providecommand \doibase [0]{http://dx.doi.org/}%
\providecommand \selectlanguage [0]{\@gobble}%
\providecommand \bibinfo  [0]{\@secondoftwo}%
\providecommand \bibfield  [0]{\@secondoftwo}%
\providecommand \translation [1]{[#1]}%
\providecommand \BibitemOpen [0]{}%
\providecommand \bibitemStop [0]{}%
\providecommand \bibitemNoStop [0]{.\EOS\space}%
\providecommand \EOS [0]{\spacefactor3000\relax}%
\providecommand \BibitemShut  [1]{\csname bibitem#1\endcsname}%
\let\auto@bib@innerbib\@empty
\bibitem [{\citenamefont {Fu}\ and\ \citenamefont {Kane}(2007)}]{Fu2007}%
  \BibitemOpen
  \bibfield  {author} {\bibinfo {author} {\bibfnamefont {L.}~\bibnamefont
  {Fu}}\ and\ \bibinfo {author} {\bibfnamefont {C.~L.}\ \bibnamefont {Kane}},\
  }\href {\doibase 10.1103/PhysRevB.76.045302} {\bibfield  {journal} {\bibinfo
  {journal} {Phys. Rev. B}\ }\textbf {\bibinfo {volume} {76}},\ \bibinfo
  {pages} {045302} (\bibinfo {year} {2007})}\BibitemShut {NoStop}%
\bibitem [{\citenamefont {Hasan}\ and\ \citenamefont {Kane}(2010)}]{Hasan2010}%
  \BibitemOpen
  \bibfield  {author} {\bibinfo {author} {\bibfnamefont {M.~Z.}\ \bibnamefont
  {Hasan}}\ and\ \bibinfo {author} {\bibfnamefont {C.~L.}\ \bibnamefont
  {Kane}},\ }\href {\doibase 10.1103/RevModPhys.82.3045} {\bibfield  {journal}
  {\bibinfo  {journal} {Rev. Mod. Phys.}\ }\textbf {\bibinfo {volume} {82}},\
  \bibinfo {pages} {3045} (\bibinfo {year} {2010})}\BibitemShut {NoStop}%
\bibitem [{\citenamefont {K\"onig}\ \emph {et~al.}(2007)\citenamefont
  {K\"onig}, \citenamefont {Wiedmann}, \citenamefont {Br\"une}, \citenamefont
  {Roth}, \citenamefont {Buhmann}, \citenamefont {Molenkamp}, \citenamefont
  {Qi},\ and\ \citenamefont {Zhang}}]{Koenig2007}%
  \BibitemOpen
  \bibfield  {author} {\bibinfo {author} {\bibfnamefont {M.}~\bibnamefont
  {K\"onig}}, \bibinfo {author} {\bibfnamefont {S.}~\bibnamefont {Wiedmann}},
  \bibinfo {author} {\bibfnamefont {C.}~\bibnamefont {Br\"une}}, \bibinfo
  {author} {\bibfnamefont {A.}~\bibnamefont {Roth}}, \bibinfo {author}
  {\bibfnamefont {H.}~\bibnamefont {Buhmann}}, \bibinfo {author} {\bibfnamefont
  {L.~W.}\ \bibnamefont {Molenkamp}}, \bibinfo {author} {\bibfnamefont {X.-L.}\
  \bibnamefont {Qi}}, \ and\ \bibinfo {author} {\bibfnamefont {S.-C.}\
  \bibnamefont {Zhang}},\ }\href
  {http://www.sciencemag.org/content/318/5851/766.abstrac} {\bibfield
  {journal} {\bibinfo  {journal} {Science}\ }\textbf {\bibinfo {volume}
  {318}},\ \bibinfo {pages} {766} (\bibinfo {year} {2007})}\BibitemShut
  {NoStop}%
\bibitem [{\citenamefont {Moore}(2010)}]{Moore2010}%
  \BibitemOpen
  \bibfield  {author} {\bibinfo {author} {\bibfnamefont {J.~E.}\ \bibnamefont
  {Moore}},\ }\href {http://dx.doi.org/10.1038/nature08916} {\bibfield
  {journal} {\bibinfo  {journal} {Nature}\ }\textbf {\bibinfo {volume} {464}},\
  \bibinfo {pages} {194} (\bibinfo {year} {2010})}\BibitemShut {NoStop}%
\bibitem [{\citenamefont {Nayak}\ \emph {et~al.}(2008)\citenamefont {Nayak},
  \citenamefont {Simon}, \citenamefont {Stern}, \citenamefont {Freedman},\ and\
  \citenamefont {Das~Sarma}}]{Nayak2008}%
  \BibitemOpen
  \bibfield  {author} {\bibinfo {author} {\bibfnamefont {C.}~\bibnamefont
  {Nayak}}, \bibinfo {author} {\bibfnamefont {S.~H.}\ \bibnamefont {Simon}},
  \bibinfo {author} {\bibfnamefont {A.}~\bibnamefont {Stern}}, \bibinfo
  {author} {\bibfnamefont {M.}~\bibnamefont {Freedman}}, \ and\ \bibinfo
  {author} {\bibfnamefont {S.}~\bibnamefont {Das~Sarma}},\ }\href
  {http://link.aps.org/doi/10.1103/RevModPhys.80.1083} {\bibfield  {journal}
  {\bibinfo  {journal} {Rev. Mod. Phys.}\ }\textbf {\bibinfo {volume} {80}},\
  \bibinfo {pages} {1083} (\bibinfo {year} {2008})}\BibitemShut {NoStop}%
\bibitem [{\citenamefont {Zutic}\ \emph {et~al.}(2004)\citenamefont {Zutic},
  \citenamefont {Fabian},\ and\ \citenamefont {Das~Sarma}}]{Zutic2004}%
  \BibitemOpen
  \bibfield  {author} {\bibinfo {author} {\bibfnamefont {I.}~\bibnamefont
  {Zutic}}, \bibinfo {author} {\bibfnamefont {J.}~\bibnamefont {Fabian}}, \
  and\ \bibinfo {author} {\bibfnamefont {S.}~\bibnamefont {Das~Sarma}},\ }\href
  {http://link.aps.org/doi/10.1103/RevModPhys.76.323} {\bibfield  {journal}
  {\bibinfo  {journal} {Rev. Mod. Phys.}\ }\textbf {\bibinfo {volume} {76}},\
  \bibinfo {pages} {323} (\bibinfo {year} {2004})}\BibitemShut {NoStop}%
\bibitem [{\citenamefont {Novoselov}\ \emph {et~al.}(2005)\citenamefont
  {Novoselov}, \citenamefont {Geim}, \citenamefont {Morozov}, \citenamefont
  {Jiang}, \citenamefont {Katsnelson}, \citenamefont {Grigorieva},
  \citenamefont {Dubonos},\ and\ \citenamefont {Firsov}}]{Novoselov2005a}%
  \BibitemOpen
  \bibfield  {author} {\bibinfo {author} {\bibfnamefont {K.~S.}\ \bibnamefont
  {Novoselov}}, \bibinfo {author} {\bibfnamefont {A.~K.}\ \bibnamefont {Geim}},
  \bibinfo {author} {\bibfnamefont {S.~V.}\ \bibnamefont {Morozov}}, \bibinfo
  {author} {\bibfnamefont {D.}~\bibnamefont {Jiang}}, \bibinfo {author}
  {\bibfnamefont {M.~I.}\ \bibnamefont {Katsnelson}}, \bibinfo {author}
  {\bibfnamefont {I.~V.}\ \bibnamefont {Grigorieva}}, \bibinfo {author}
  {\bibfnamefont {S.~V.}\ \bibnamefont {Dubonos}}, \ and\ \bibinfo {author}
  {\bibfnamefont {A.~A.}\ \bibnamefont {Firsov}},\ }\href
  {http://dx.doi.org/10.1038/nature04233} {\bibfield  {journal} {\bibinfo
  {journal} {Nature}\ }\textbf {\bibinfo {volume} {438}},\ \bibinfo {pages}
  {197} (\bibinfo {year} {2005})}\BibitemShut {NoStop}%
\bibitem [{\citenamefont {Zhang}\ \emph {et~al.}(2005)\citenamefont {Zhang},
  \citenamefont {Tan}, \citenamefont {Stormer},\ and\ \citenamefont
  {Kim}}]{Zhang2005}%
  \BibitemOpen
  \bibfield  {author} {\bibinfo {author} {\bibfnamefont {Y.}~\bibnamefont
  {Zhang}}, \bibinfo {author} {\bibfnamefont {Y.-W.}\ \bibnamefont {Tan}},
  \bibinfo {author} {\bibfnamefont {H.~L.}\ \bibnamefont {Stormer}}, \ and\
  \bibinfo {author} {\bibfnamefont {P.}~\bibnamefont {Kim}},\ }\href
  {http://dx.doi.org/10.1038/nature04235} {\bibfield  {journal} {\bibinfo
  {journal} {Nature}\ }\textbf {\bibinfo {volume} {438}},\ \bibinfo {pages}
  {201} (\bibinfo {year} {2005})}\BibitemShut {NoStop}%
\bibitem [{\citenamefont {Geim}(2009)}]{Geim2009}%
  \BibitemOpen
  \bibfield  {author} {\bibinfo {author} {\bibfnamefont {A.~K.}\ \bibnamefont
  {Geim}},\ }\href {\doibase 10.1126/science.1158877} {\bibfield  {journal}
  {\bibinfo  {journal} {Science}\ }\textbf {\bibinfo {volume} {324}},\ \bibinfo
  {pages} {1530} (\bibinfo {year} {2009})}\BibitemShut {NoStop}%
\bibitem [{\citenamefont {Castro~Neto}\ \emph {et~al.}(2009)\citenamefont
  {Castro~Neto}, \citenamefont {Guinea}, \citenamefont {Peres}, \citenamefont
  {Novoselov},\ and\ \citenamefont {Geim}}]{CastroNeto2009}%
  \BibitemOpen
  \bibfield  {author} {\bibinfo {author} {\bibfnamefont {A.~H.}\ \bibnamefont
  {Castro~Neto}}, \bibinfo {author} {\bibfnamefont {F.}~\bibnamefont {Guinea}},
  \bibinfo {author} {\bibfnamefont {N.~M.~R.}\ \bibnamefont {Peres}}, \bibinfo
  {author} {\bibfnamefont {K.~S.}\ \bibnamefont {Novoselov}}, \ and\ \bibinfo
  {author} {\bibfnamefont {A.~K.}\ \bibnamefont {Geim}},\ }\href {\doibase
  10.1103/RevModPhys.81.109} {\bibfield  {journal} {\bibinfo  {journal} {Rev.
  Mod. Phys.}\ }\textbf {\bibinfo {volume} {81}},\ \bibinfo {pages} {109}
  (\bibinfo {year} {2009})}\BibitemShut {NoStop}%
\bibitem [{\citenamefont {Balandin}\ \emph {et~al.}(2008)\citenamefont
  {Balandin}, \citenamefont {Ghosh}, \citenamefont {Bao}, \citenamefont
  {Calizo}, \citenamefont {Teweldebrhan}, \citenamefont {Miao},\ and\
  \citenamefont {Lau}}]{Balandin2008}%
  \BibitemOpen
  \bibfield  {author} {\bibinfo {author} {\bibfnamefont {A.~A.}\ \bibnamefont
  {Balandin}}, \bibinfo {author} {\bibfnamefont {S.}~\bibnamefont {Ghosh}},
  \bibinfo {author} {\bibfnamefont {W.}~\bibnamefont {Bao}}, \bibinfo {author}
  {\bibfnamefont {I.}~\bibnamefont {Calizo}}, \bibinfo {author} {\bibfnamefont
  {D.}~\bibnamefont {Teweldebrhan}}, \bibinfo {author} {\bibfnamefont
  {F.}~\bibnamefont {Miao}}, \ and\ \bibinfo {author} {\bibfnamefont {C.~N.}\
  \bibnamefont {Lau}},\ }\bibfield  {booktitle} {\emph {\bibinfo {booktitle}
  {Nano Letters}},\ }\href {\doibase 10.1021/nl0731872} {\bibfield  {journal}
  {\bibinfo  {journal} {Nano Lett.}\ }\textbf {\bibinfo {volume} {8}},\
  \bibinfo {pages} {902} (\bibinfo {year} {2008})}\BibitemShut {NoStop}%
\bibitem [{\citenamefont {Bonaccorso}\ \emph {et~al.}(2010)\citenamefont
  {Bonaccorso}, \citenamefont {Sun}, \citenamefont {Hasan},\ and\ \citenamefont
  {Ferrari}}]{Bonaccorso2010}%
  \BibitemOpen
  \bibfield  {author} {\bibinfo {author} {\bibfnamefont {F.}~\bibnamefont
  {Bonaccorso}}, \bibinfo {author} {\bibfnamefont {Z.}~\bibnamefont {Sun}},
  \bibinfo {author} {\bibfnamefont {T.}~\bibnamefont {Hasan}}, \ and\ \bibinfo
  {author} {\bibfnamefont {A.~C.}\ \bibnamefont {Ferrari}},\ }\href
  {http://dx.doi.org/10.1038/nphoton.2010.186} {\bibfield  {journal} {\bibinfo
  {journal} {Nat Photon}\ }\textbf {\bibinfo {volume} {4}},\ \bibinfo {pages}
  {611} (\bibinfo {year} {2010})}\BibitemShut {NoStop}%
\bibitem [{\citenamefont {Xia}\ \emph {et~al.}(2009)\citenamefont {Xia},
  \citenamefont {Mueller}, \citenamefont {Lin}, \citenamefont {Valdes-Garcia},\
  and\ \citenamefont {Avouris}}]{Xia2009}%
  \BibitemOpen
  \bibfield  {author} {\bibinfo {author} {\bibfnamefont {F.}~\bibnamefont
  {Xia}}, \bibinfo {author} {\bibfnamefont {T.}~\bibnamefont {Mueller}},
  \bibinfo {author} {\bibfnamefont {Y.-m.}\ \bibnamefont {Lin}}, \bibinfo
  {author} {\bibfnamefont {A.}~\bibnamefont {Valdes-Garcia}}, \ and\ \bibinfo
  {author} {\bibfnamefont {P.}~\bibnamefont {Avouris}},\ }\href
  {http://dx.doi.org/10.1038/nnano.2009.292} {\bibfield  {journal} {\bibinfo
  {journal} {Nat Nano}\ }\textbf {\bibinfo {volume} {4}},\ \bibinfo {pages}
  {839} (\bibinfo {year} {2009})}\BibitemShut {NoStop}%
\bibitem [{\citenamefont {Ren}\ \emph {et~al.}(2012)\citenamefont {Ren},
  \citenamefont {Zhang}, \citenamefont {Yao}, \citenamefont {Sun},
  \citenamefont {Kaneko}, \citenamefont {Yan}, \citenamefont {Nanot},
  \citenamefont {Jin}, \citenamefont {Kawayama}, \citenamefont {Tonouchi},
  \citenamefont {Tour},\ and\ \citenamefont {Kono}}]{Ren2012}%
  \BibitemOpen
  \bibfield  {author} {\bibinfo {author} {\bibfnamefont {L.}~\bibnamefont
  {Ren}}, \bibinfo {author} {\bibfnamefont {Q.}~\bibnamefont {Zhang}}, \bibinfo
  {author} {\bibfnamefont {J.}~\bibnamefont {Yao}}, \bibinfo {author}
  {\bibfnamefont {Z.}~\bibnamefont {Sun}}, \bibinfo {author} {\bibfnamefont
  {R.}~\bibnamefont {Kaneko}}, \bibinfo {author} {\bibfnamefont
  {Z.}~\bibnamefont {Yan}}, \bibinfo {author} {\bibfnamefont {S.}~\bibnamefont
  {Nanot}}, \bibinfo {author} {\bibfnamefont {Z.}~\bibnamefont {Jin}}, \bibinfo
  {author} {\bibfnamefont {I.}~\bibnamefont {Kawayama}}, \bibinfo {author}
  {\bibfnamefont {M.}~\bibnamefont {Tonouchi}}, \bibinfo {author}
  {\bibfnamefont {J.~M.}\ \bibnamefont {Tour}}, \ and\ \bibinfo {author}
  {\bibfnamefont {J.}~\bibnamefont {Kono}},\ }\bibfield  {booktitle} {\emph
  {\bibinfo {booktitle} {Nano Letters}},\ }\href {\doibase 10.1021/nl301496r}
  {\bibfield  {journal} {\bibinfo  {journal} {Nano Lett.}\ }\textbf {\bibinfo
  {volume} {12}},\ \bibinfo {pages} {3711} (\bibinfo {year}
  {2012})}\BibitemShut {NoStop}%
\bibitem [{\citenamefont {Tielrooij}\ \emph {et~al.}(2013)\citenamefont
  {Tielrooij}, \citenamefont {Song}, \citenamefont {Jensen}, \citenamefont
  {Centeno}, \citenamefont {Pesquera}, \citenamefont {Zurutuza~Elorza},
  \citenamefont {Bonn}, \citenamefont {Levitov},\ and\ \citenamefont
  {Koppens}}]{Tielrooij2013}%
  \BibitemOpen
  \bibfield  {author} {\bibinfo {author} {\bibfnamefont {K.~J.}\ \bibnamefont
  {Tielrooij}}, \bibinfo {author} {\bibfnamefont {J.~C.~W.}\ \bibnamefont
  {Song}}, \bibinfo {author} {\bibfnamefont {S.~A.}\ \bibnamefont {Jensen}},
  \bibinfo {author} {\bibfnamefont {A.}~\bibnamefont {Centeno}}, \bibinfo
  {author} {\bibfnamefont {A.}~\bibnamefont {Pesquera}}, \bibinfo {author}
  {\bibfnamefont {A.}~\bibnamefont {Zurutuza~Elorza}}, \bibinfo {author}
  {\bibfnamefont {M.}~\bibnamefont {Bonn}}, \bibinfo {author} {\bibfnamefont
  {L.~S.}\ \bibnamefont {Levitov}}, \ and\ \bibinfo {author} {\bibfnamefont
  {F.~H.~L.}\ \bibnamefont {Koppens}},\ }\href
  {http://dx.doi.org/10.1038/nphys2564} {\bibfield  {journal} {\bibinfo
  {journal} {Nat Phys}\ }\textbf {\bibinfo {volume} {9}},\ \bibinfo {pages}
  {248} (\bibinfo {year} {2013})}\BibitemShut {NoStop}%
\bibitem [{\citenamefont {Oka}\ and\ \citenamefont {Aoki}(2009)}]{Oka2009}%
  \BibitemOpen
  \bibfield  {author} {\bibinfo {author} {\bibfnamefont {T.}~\bibnamefont
  {Oka}}\ and\ \bibinfo {author} {\bibfnamefont {H.}~\bibnamefont {Aoki}},\
  }\href {http://link.aps.org/doi/10.1103/PhysRevB.79.081406} {\bibfield
  {journal} {\bibinfo  {journal} {Phys. Rev. B}\ }\textbf {\bibinfo {volume}
  {79}},\ \bibinfo {pages} {081406} (\bibinfo {year} {2009})}\BibitemShut
  {NoStop}%
\bibitem [{\citenamefont {L\'opez-Rodr\'{\i}guez}\ and\ \citenamefont
  {Naumis}(2008)}]{Lopez-Rodriguez2008}%
  \BibitemOpen
  \bibfield  {author} {\bibinfo {author} {\bibfnamefont {F.~J.}\ \bibnamefont
  {L\'opez-Rodr\'{\i}guez}}\ and\ \bibinfo {author} {\bibfnamefont {G.~G.}\
  \bibnamefont {Naumis}},\ }\href
  {http://link.aps.org/doi/10.1103/PhysRevB.78.201406} {\bibfield  {journal}
  {\bibinfo  {journal} {Phys. Rev. B}\ }\textbf {\bibinfo {volume} {78}},\
  \bibinfo {pages} {201406} (\bibinfo {year} {2008})}\BibitemShut {NoStop}%
\bibitem [{\citenamefont {Kibis}(2010)}]{Kibis2010}%
  \BibitemOpen
  \bibfield  {author} {\bibinfo {author} {\bibfnamefont {O.~V.}\ \bibnamefont
  {Kibis}},\ }\href {http://link.aps.org/doi/10.1103/PhysRevB.81.165433}
  {\bibfield  {journal} {\bibinfo  {journal} {Phys. Rev. B}\ }\textbf {\bibinfo
  {volume} {81}},\ \bibinfo {pages} {165433} (\bibinfo {year}
  {2010})}\BibitemShut {NoStop}%
\bibitem [{\citenamefont {Calvo}\ \emph {et~al.}(2011)\citenamefont {Calvo},
  \citenamefont {Pastawski}, \citenamefont {Roche},\ and\ \citenamefont
  {Foa~Torres}}]{Calvo2011}%
  \BibitemOpen
  \bibfield  {author} {\bibinfo {author} {\bibfnamefont {H.~L.}\ \bibnamefont
  {Calvo}}, \bibinfo {author} {\bibfnamefont {H.~M.}\ \bibnamefont
  {Pastawski}}, \bibinfo {author} {\bibfnamefont {S.}~\bibnamefont {Roche}}, \
  and\ \bibinfo {author} {\bibfnamefont {L.~E.~F.}\ \bibnamefont
  {Foa~Torres}},\ }\href {http://dx.doi.org/10.1063/1.3597412} {\bibfield
  {journal} {\bibinfo  {journal} {Appl. Phys. Lett.}\ }\textbf {\bibinfo
  {volume} {98}},\ \bibinfo {pages} {232103} (\bibinfo {year}
  {2011})}\BibitemShut {NoStop}%
\bibitem [{\citenamefont {Savel'ev}\ and\ \citenamefont
  {Alexandrov}(2011)}]{Savelev2011}%
  \BibitemOpen
  \bibfield  {author} {\bibinfo {author} {\bibfnamefont {S.~E.}\ \bibnamefont
  {Savel'ev}}\ and\ \bibinfo {author} {\bibfnamefont {A.~S.}\ \bibnamefont
  {Alexandrov}},\ }\href {http://link.aps.org/doi/10.1103/PhysRevB.84.035428}
  {\bibfield  {journal} {\bibinfo  {journal} {Phys. Rev. B}\ }\textbf {\bibinfo
  {volume} {84}},\ \bibinfo {pages} {035428} (\bibinfo {year}
  {2011})}\BibitemShut {NoStop}%
\bibitem [{\citenamefont {Wang}\ \emph {et~al.}(2013)\citenamefont {Wang},
  \citenamefont {Steinberg}, \citenamefont {Jarillo-Herrero},\ and\
  \citenamefont {Gedik}}]{Wang2013}%
  \BibitemOpen
  \bibfield  {author} {\bibinfo {author} {\bibfnamefont {Y.~H.}\ \bibnamefont
  {Wang}}, \bibinfo {author} {\bibfnamefont {H.}~\bibnamefont {Steinberg}},
  \bibinfo {author} {\bibfnamefont {P.}~\bibnamefont {Jarillo-Herrero}}, \ and\
  \bibinfo {author} {\bibfnamefont {N.}~\bibnamefont {Gedik}},\ }\href
  {\doibase 10.1126/science.1239834} {\bibfield  {journal} {\bibinfo  {journal}
  {Science}\ }\textbf {\bibinfo {volume} {342}},\ \bibinfo {pages} {453}
  (\bibinfo {year} {2013})}\BibitemShut {NoStop}%
\bibitem [{not({\natexlab{a}})}]{notes1}%
  \BibitemOpen
  \href@noop {} {} \bibinfo {note} {The mini-gap at the
  Dirac point involves a virtual process with photon emission and reabsorption.
  Previous studies inferred on the topological properties at this mini-gap
  based on calculations in the bulk \cite{Kitagawa2011,SuarezMorell2012} or for
  ultrasmall ribbons \cite{Gu2011}}\BibitemShut {NoStop}%
\bibitem [{\citenamefont {Syzranov}\ \emph {et~al.}(2008)\citenamefont
  {Syzranov}, \citenamefont {Fistul},\ and\ \citenamefont
  {Efetov}}]{Syzranov2008}%
  \BibitemOpen
  \bibfield  {author} {\bibinfo {author} {\bibfnamefont {S.~V.}\ \bibnamefont
  {Syzranov}}, \bibinfo {author} {\bibfnamefont {M.~V.}\ \bibnamefont
  {Fistul}}, \ and\ \bibinfo {author} {\bibfnamefont {K.~B.}\ \bibnamefont
  {Efetov}},\ }\href {http://link.aps.org/doi/10.1103/PhysRevB.78.045407}
  {\bibfield  {journal} {\bibinfo  {journal} {Phys. Rev. B}\ }\textbf {\bibinfo
  {volume} {78}},\ \bibinfo {pages} {045407} (\bibinfo {year}
  {2008})}\BibitemShut {NoStop}%
\bibitem [{\citenamefont {Lindner}\ \emph {et~al.}(2011)\citenamefont
  {Lindner}, \citenamefont {Refael},\ and\ \citenamefont
  {Galitski}}]{Lindner2011}%
  \BibitemOpen
  \bibfield  {author} {\bibinfo {author} {\bibfnamefont {N.~H.}\ \bibnamefont
  {Lindner}}, \bibinfo {author} {\bibfnamefont {G.}~\bibnamefont {Refael}}, \
  and\ \bibinfo {author} {\bibfnamefont {V.}~\bibnamefont {Galitski}},\ }\href
  {http://dx.doi.org/10.1038/nphys1926} {\bibfield  {journal} {\bibinfo
  {journal} {Nat Phys}\ }\textbf {\bibinfo {volume} {7}},\ \bibinfo {pages}
  {490} (\bibinfo {year} {2011})}\BibitemShut {NoStop}%
\bibitem [{\citenamefont {Kitagawa}\ \emph {et~al.}(2010)\citenamefont
  {Kitagawa}, \citenamefont {Berg}, \citenamefont {Rudner},\ and\ \citenamefont
  {Demler}}]{Kitagawa2010}%
  \BibitemOpen
  \bibfield  {author} {\bibinfo {author} {\bibfnamefont {T.}~\bibnamefont
  {Kitagawa}}, \bibinfo {author} {\bibfnamefont {E.}~\bibnamefont {Berg}},
  \bibinfo {author} {\bibfnamefont {M.}~\bibnamefont {Rudner}}, \ and\ \bibinfo
  {author} {\bibfnamefont {E.}~\bibnamefont {Demler}},\ }\href
  {http://link.aps.org/doi/10.1103/PhysRevB.82.235114} {\bibfield  {journal}
  {\bibinfo  {journal} {Phys. Rev. B}\ }\textbf {\bibinfo {volume} {82}},\
  \bibinfo {pages} {235114} (\bibinfo {year} {2010})}\BibitemShut {NoStop}%
\bibitem [{oth()}]{other}%
  \BibitemOpen
  \href@noop {} {}\bibinfo {note} {Other recent related studies include: Y.
  Tenenbaum Katan and D. Podolsky, arXiv:1309.0203 [cond-mat.str-el];
  Gomez-Leon et al. arXiv:1309.5402 [cond-mat.mes-hall]}\BibitemShut {NoStop}%
\bibitem [{\citenamefont {Rechtsman}\ \emph {et~al.}(2013)\citenamefont
  {Rechtsman}, \citenamefont {Zeuner}, \citenamefont {Plotnik}, \citenamefont
  {Lumer}, \citenamefont {Podolsky}, \citenamefont {Dreisow}, \citenamefont
  {Nolte}, \citenamefont {Segev},\ and\ \citenamefont
  {Szameit}}]{Rechtsman2013}%
  \BibitemOpen
  \bibfield  {author} {\bibinfo {author} {\bibfnamefont {M.~C.}\ \bibnamefont
  {Rechtsman}}, \bibinfo {author} {\bibfnamefont {J.~M.}\ \bibnamefont
  {Zeuner}}, \bibinfo {author} {\bibfnamefont {Y.}~\bibnamefont {Plotnik}},
  \bibinfo {author} {\bibfnamefont {Y.}~\bibnamefont {Lumer}}, \bibinfo
  {author} {\bibfnamefont {D.}~\bibnamefont {Podolsky}}, \bibinfo {author}
  {\bibfnamefont {F.}~\bibnamefont {Dreisow}}, \bibinfo {author} {\bibfnamefont
  {S.}~\bibnamefont {Nolte}}, \bibinfo {author} {\bibfnamefont
  {M.}~\bibnamefont {Segev}}, \ and\ \bibinfo {author} {\bibfnamefont
  {A.}~\bibnamefont {Szameit}},\ }\href {http://dx.doi.org/10.1038/nature12066}
  {\bibfield  {journal} {\bibinfo  {journal} {Nature}\ }\textbf {\bibinfo
  {volume} {496}},\ \bibinfo {pages} {196} (\bibinfo {year}
  {2013})}\BibitemShut {NoStop}%
\bibitem [{\citenamefont {Cayssol}\ \emph {et~al.}(2013)\citenamefont
  {Cayssol}, \citenamefont {D\'{o}ra}, \citenamefont {Simon},\ and\
  \citenamefont {Moessner}}]{Cayssol2013}%
  \BibitemOpen
  \bibfield  {author} {\bibinfo {author} {\bibfnamefont {J.}~\bibnamefont
  {Cayssol}}, \bibinfo {author} {\bibfnamefont {B.}~\bibnamefont {D\'{o}ra}},
  \bibinfo {author} {\bibfnamefont {F.}~\bibnamefont {Simon}}, \ and\ \bibinfo
  {author} {\bibfnamefont {R.}~\bibnamefont {Moessner}},\ }\href
  {http://dx.doi.org/10.1002/pssr.201206451} {\bibfield  {journal} {\bibinfo
  {journal} {Phys. Status Solidi RRL}\ }\textbf {\bibinfo {volume} {7}},\
  \bibinfo {pages} {101} (\bibinfo {year} {2013})}\BibitemShut {NoStop}%
\bibitem [{\citenamefont {Glazov}\ and\ \citenamefont
  {Ganichev}(2013)}]{Glazov2013}%
  \BibitemOpen
  \bibfield  {author} {\bibinfo {author} {\bibfnamefont {M.}~\bibnamefont
  {Glazov}}\ and\ \bibinfo {author} {\bibfnamefont {S.}~\bibnamefont
  {Ganichev}},\ }\href@noop {} {\bibfield  {journal} {\bibinfo  {journal}
  {arXiv:1306.2049 [cond-mat.mes-hall]}\ ,\ \bibinfo {pages} {unpublished}}
  (\bibinfo {year} {2013})}\BibitemShut {NoStop}%
\bibitem [{\citenamefont {Shirley}(1965)}]{Shirley1965}%
  \BibitemOpen
  \bibfield  {author} {\bibinfo {author} {\bibfnamefont {J.~H.}\ \bibnamefont
  {Shirley}},\ }\href {\doibase 10.1103/PhysRev.138.B979} {\bibfield  {journal}
  {\bibinfo  {journal} {Phys. Rev.}\ }\textbf {\bibinfo {volume} {138}},\
  \bibinfo {pages} {B979} (\bibinfo {year} {1965})}\BibitemShut {NoStop}%
\bibitem [{\citenamefont {Grifoni}\ and\ \citenamefont
  {H\"anggi}(1998)}]{Grifoni1998}%
  \BibitemOpen
  \bibfield  {author} {\bibinfo {author} {\bibfnamefont {M.}~\bibnamefont
  {Grifoni}}\ and\ \bibinfo {author} {\bibfnamefont {P.}~\bibnamefont
  {H\"anggi}},\ }\href {\doibase
  http://dx.doi.org/10.1016/S0370-1573(98)00022-2} {\bibfield  {journal}
  {\bibinfo  {journal} {Physics Reports}\ }\textbf {\bibinfo {volume} {304}},\
  \bibinfo {pages} {229 } (\bibinfo {year} {1998})}\BibitemShut {NoStop}%
\bibitem [{\citenamefont {Kohler}\ \emph {et~al.}(2005)\citenamefont {Kohler},
  \citenamefont {Lehmann},\ and\ \citenamefont {H\"anggi}}]{Kohler2005}%
  \BibitemOpen
  \bibfield  {author} {\bibinfo {author} {\bibfnamefont {S.}~\bibnamefont
  {Kohler}}, \bibinfo {author} {\bibfnamefont {J.}~\bibnamefont {Lehmann}}, \
  and\ \bibinfo {author} {\bibfnamefont {P.}~\bibnamefont {H\"anggi}},\ }\href
  {\doibase 10.1016/j.physrep.2004.11.002} {\bibfield  {journal} {\bibinfo
  {journal} {Physics Reports}\ }\textbf {\bibinfo {volume} {406}},\ \bibinfo
  {pages} {379} (\bibinfo {year} {2005})}\BibitemShut {NoStop}%
\bibitem [{\citenamefont {Rudner}\ \emph {et~al.}(2013)\citenamefont {Rudner},
  \citenamefont {Lindner}, \citenamefont {Berg},\ and\ \citenamefont
  {Levin}}]{Rudner2013}%
  \BibitemOpen
  \bibfield  {author} {\bibinfo {author} {\bibfnamefont {M.~S.}\ \bibnamefont
  {Rudner}}, \bibinfo {author} {\bibfnamefont {N.~H.}\ \bibnamefont {Lindner}},
  \bibinfo {author} {\bibfnamefont {E.}~\bibnamefont {Berg}}, \ and\ \bibinfo
  {author} {\bibfnamefont {M.}~\bibnamefont {Levin}},\ }\href@noop {}
  {\bibfield  {journal} {\bibinfo  {journal} {Phys. Rev. X}\ }\textbf {\bibinfo
  {volume} {3}},\ \bibinfo {pages} {031005} (\bibinfo {year} {2013})},\
  \bibinfo {note} {arXiv:1212.3324 [cond-mat.mes-hall]}\BibitemShut {NoStop}%
\bibitem [{\citenamefont {Brey}\ and\ \citenamefont {Fertig}(2006)}]{Brey2006}%
  \BibitemOpen
  \bibfield  {author} {\bibinfo {author} {\bibfnamefont {L.}~\bibnamefont
  {Brey}}\ and\ \bibinfo {author} {\bibfnamefont {H.~A.}\ \bibnamefont
  {Fertig}},\ }\href {\doibase 10.1103/PhysRevB.73.235411} {\bibfield
  {journal} {\bibinfo  {journal} {Phys. Rev. B}\ }\textbf {\bibinfo {volume}
  {73}},\ \bibinfo {pages} {235411} (\bibinfo {year} {2006})}\BibitemShut
  {NoStop}%
\bibitem [{\citenamefont {Usaj}(2009)}]{Usaj2009}%
  \BibitemOpen
  \bibfield  {author} {\bibinfo {author} {\bibfnamefont {G.}~\bibnamefont
  {Usaj}},\ }\href {\doibase 10.1103/PhysRevB.80.081414} {\bibfield  {journal}
  {\bibinfo  {journal} {Phys. Rev. B}\ }\textbf {\bibinfo {volume} {80}},\
  \bibinfo {pages} {081414} (\bibinfo {year} {2009})}\BibitemShut {NoStop}%
\bibitem [{not({\natexlab{b}})}]{notes2}%
  \BibitemOpen
  \href@noop {} {} \bibinfo {note} {The same result can be
  obtained by explicitly calculating the average value of the velocity operator
  $v_F\langle\langle\sigma_y\rangle\rangle$, where
  $\langle\langle\dots\rangle\rangle$ indicates averaging on a time
  period}\BibitemShut {NoStop}%
\bibitem [{not({\natexlab{c}})}]{note-transport}%
  \BibitemOpen
  \href@noop {} {} \bibinfo {note} {In the transport setup
  illumination is assumed to be limited to the sample region only. Dissipation
  of the excess energy is then assumed to take place in the leads, allowing for
  a scattering formulation of the transport problem \cite{Kohler2005}. See also
  the supplementary information}\BibitemShut {NoStop}%
\bibitem [{\citenamefont {Kitagawa}\ \emph {et~al.}(2011)\citenamefont
  {Kitagawa}, \citenamefont {Oka}, \citenamefont {Brataas}, \citenamefont
  {Fu},\ and\ \citenamefont {Demler}}]{Kitagawa2011}%
  \BibitemOpen
  \bibfield  {author} {\bibinfo {author} {\bibfnamefont {T.}~\bibnamefont
  {Kitagawa}}, \bibinfo {author} {\bibfnamefont {T.}~\bibnamefont {Oka}},
  \bibinfo {author} {\bibfnamefont {A.}~\bibnamefont {Brataas}}, \bibinfo
  {author} {\bibfnamefont {L.}~\bibnamefont {Fu}}, \ and\ \bibinfo {author}
  {\bibfnamefont {E.}~\bibnamefont {Demler}},\ }\href
  {http://link.aps.org/doi/10.1103/PhysRevB.84.235108} {\bibfield  {journal}
  {\bibinfo  {journal} {Phys. Rev. B}\ }\textbf {\bibinfo {volume} {84}},\
  \bibinfo {pages} {235108} (\bibinfo {year} {2011})}\BibitemShut {NoStop}%
\bibitem [{\citenamefont {Suarez~Morell}\ and\ \citenamefont
  {Foa~Torres}(2012)}]{SuarezMorell2012}%
  \BibitemOpen
  \bibfield  {author} {\bibinfo {author} {\bibfnamefont {E.}~\bibnamefont
  {Suarez~Morell}}\ and\ \bibinfo {author} {\bibfnamefont {L.~E.~F.}\
  \bibnamefont {Foa~Torres}},\ }\href@noop {} {\bibfield  {journal} {\bibinfo
  {journal} {Physical Review B}\ }\textbf {\bibinfo {volume} {86}},\ \bibinfo
  {pages} {125449} (\bibinfo {year} {2012})}\BibitemShut {NoStop}%
\bibitem [{\citenamefont {Gu}\ \emph {et~al.}(2011)\citenamefont {Gu},
  \citenamefont {Fertig}, \citenamefont {Arovas},\ and\ \citenamefont
  {Auerbach}}]{Gu2011}%
  \BibitemOpen
  \bibfield  {author} {\bibinfo {author} {\bibfnamefont {Z.}~\bibnamefont
  {Gu}}, \bibinfo {author} {\bibfnamefont {H.~A.}\ \bibnamefont {Fertig}},
  \bibinfo {author} {\bibfnamefont {D.~P.}\ \bibnamefont {Arovas}}, \ and\
  \bibinfo {author} {\bibfnamefont {A.}~\bibnamefont {Auerbach}},\ }\href
  {http://link.aps.org/doi/10.1103/PhysRevLett.107.216601} {\bibfield
  {journal} {\bibinfo  {journal} {Phys. Rev. Lett.}\ }\textbf {\bibinfo
  {volume} {107}},\ \bibinfo {pages} {216601} (\bibinfo {year}
  {2011})}\BibitemShut {NoStop}%
\end{thebibliography}
\end{document}